\documentclass[11pt]{article} 
\usepackage{moriond2000,epsfig}

\def\be{\begin{equation}} 
\def\ee{\end{equation}} 
\def\bea{\begin{eqnarray}} 
\def\eea{\end{eqnarray}} 
 
\begin{document} 
\vspace*{4cm} 
\title{CLUSTER GALAXIES: CONTRIBUTION TO THE ARC STATISTICS} 
 
\author{ M. MENEGHETTI$^{1,2}$, M. BOLZONELLA$^3$, M. BARTELMANN$^2$, \\ 
L. MOSCARDINI$^1$ AND G. TORMEN$^1$} 
 
\address{$^1$Dipartimento di Astronomia, Universit\`a di Padova, vicolo 
dell'Osservatorio 5, I-35122 Padova, Italy \\ 
$^2$Max Planck Institut f\"ur Astrophysik, P.O. Box 1523, D-85740 
Garching, Germany\\ 
$^3$Istituto di Fisica Cosmica G.P.S. Occhialini, via Bassini 15, I-20133 
Milano, Italy} 
 
\maketitle\abstracts{ 
We present the results of a set of numerical simulations aiming at 
evaluating the effects of cluster galaxies on the arc statistics.  At 
this goal we use nine different galaxy clusters obtained from N-body 
simulations. We mimic the presence of a population of galaxies inside 
each cluster, trying to reproduce the observed luminosity function and 
the spatial distribution.  We compare the statistical distributions of 
the properties of the gravitational arcs produced by our clusters with 
and without galaxies.  We find that the cluster galaxies do not 
introduce perturbations strong enough to significantly change the 
number of arcs and the distributions of the properties of long arcs.} 
 
\section{Introduction} 
 
The statistics of arcs is a potentially very sensitive probe for the 
cosmological matter density parameter $\Omega_0$ and for 
$\Omega_\Lambda$. Theoretical studies showed that more arcs can be 
expected in a universe with low density and small cosmological 
constant. In fact, low density makes clusters form earlier, and a low 
cosmological constant makes them more compact individually.  Previous 
analyses (e.g. Bartelmann et al.~1998) neglected the granularity of 
the gravitational cluster potentials due to the presence of galaxies 
which can be considered using numerical simulations. 
 
\section{The cluster sample} 
 
The simulated clusters used as lenses in the present analysis are 
presented in Tormen et al. \cite{tormen97}.  The sample is formed by 
the nine most massive clusters obtained in a cosmological simulation 
of an Einstein-de Sitter universe, evolved using a P3M code.  The 
original box-size is $L=150$ Mpc (a Hubble constant of 50 km s$^{-1}$ 
Mpc$^{-1}$ is used).  The initial conditions have a scale-free power 
spectrum $P(k) \propto k^{-1}$, very close to the behaviour of the 
standard CDM model on the scales relevant for cluster formation. The 
normalization is chosen to roughly match the observed local abundance 
of clusters \cite{white93} and corresponds to $\sigma_8=0.63$, where 
$\sigma_8$ is the r.m.s. matter density fluctuation in spheres of 
radius $r=8 h^{-1}$ Mpc.  Each cluster was obtained using a tree/SPH 
code adopting a re-simulation technique \cite{tormen97}, which allows 
a much higher spatial and mass resolution.  The masses of these 
clusters are in the range $5 \times 10^{14} \div 3 
\times 10^{15} M_{\odot}$. 
 
\section{Lensing properties of the clusters} 
We centred each cluster in a cube of side $6$ Mpc, where we placed a 
regular grid of $N_g=128^3$ cells.  For each of them we extracted 
three different surface-density fields $\Sigma$, by projecting the 3-D 
density field $\rho$ (obtained by {\em Triangular Shape Cloud} method) 
along the three cartesian axes.  This produces three lens planes, 
which we consider as independent cluster models for the purpose of 
this study.  This made possible to perform 27 different lensing 
simulations starting from our sample of 9 clusters.  We fix the 
redshifts of the lens and of the source planes equal to $z_L=0.4$ and 
$z_S=2$ respectively, leading to a value of the critical density for 
our clusters of $\Sigma_{cr} \simeq 1.975 \times 10^{15} 
M_{\odot}/$Mpc$^2$.  We shot a bundle of $1024 \times 1024$ rays 
across the central fourth of the lens plane, where most of the cluster 
mass is projected. In fact, our goal is to study the strong lensing 
properties of the clusters.  The deflection angle $\alpha$ of each ray 
is computed by summing the contribution from all the cells of the grid 
on which $\Sigma$ is defined. Finally, solving the lens equation, the 
arrival positions of the rays on the source plane are calculated. 
Once the deflection angles are known, all the lensing properties of 
the cluster can be easily evaluated. 
 
\section{Simulating the galaxy distribution inside the 
cluster} To simulate a population of galaxy lenses inside the cluster, 
in such a way that their observational properties are well reproduced, 
we start from the luminosity function of the Coma cluster, whose mass 
is similar to that or our simulated clusters. This luminosity function 
has been recently derived in the V-band by Lobo et 
al. \cite{lobo97}. In the magnitude range $13.5 < V \leq 21.0$ 
(corresponding to the absolute magnitude range $-22.24 < M_{V} \leq 
-14.74$) it is well described by the combination of a steep Schechter 
function and of a Gaussian function. 
 
Using Monte Carlo methods, we generate a sample of galaxies with 
luminosities distributed in a way close to the Coma cluster 
galaxies. To convert the luminosity to masses, we take the average 
relation $\langle M/L \rangle = 3.2 (M_{\odot}/L_{\odot})$ (see 
\cite{white93,vandermarel91}). In this procedure the total number of 
galaxies to place into each simulated cluster is determined by 
imposing a baryonic fraction equal to that estimated by White et 
al. \cite{white93} for the Coma cluster ($M_b/M_{tot} \simeq 0.009$, 
where $M_b$ is the baryonic mass in galaxies and $M_{tot}$ the total 
mass within the Abell radius). Moreover, to consider the presence of a 
dark matter halo around each galaxy, we obtain the total (virial) 
masses $M_{vir}$ by multiplying the baryonic masses previously 
obtained by the factor $f_b^{-1}$, where $f_b$ represents the average 
baryonic fraction inside single galaxies. As this quantity is not well 
known observationally, we take a fiducial value $f_b$ ($\sim 5 \%$), 
close to the value predicted by the standard model of primordial 
nucleosynthesis.  To place galaxies inside the cluster in a realistic 
way, we made the assumptions that the galaxy number density should 
follow the total density field and that the most massive galaxies 
should be placed at the centre of the cluster or in other large 
subclumps. 
 
The galaxies are modelled as spheres with a NFW density profile 
\cite{nfw97}, truncated at a cut-off radius, where the galaxy density 
falls below the local cluster density. In fact, for lensing analysis 
we are interested only to the galaxy mass which emerges from the mean 
local density of the cluster.  Galaxies placed close to the cluster 
centre have a smaller radius because the cluster density is higher 
there, so only a small part of the galaxy profile can emerge.  The 
galaxy contribution to the deflection angles can be analytically 
computed and summed to the contribution from the remaining dark matter 
in the cluster. 
 
\begin{figure} 
{\psfig{figure=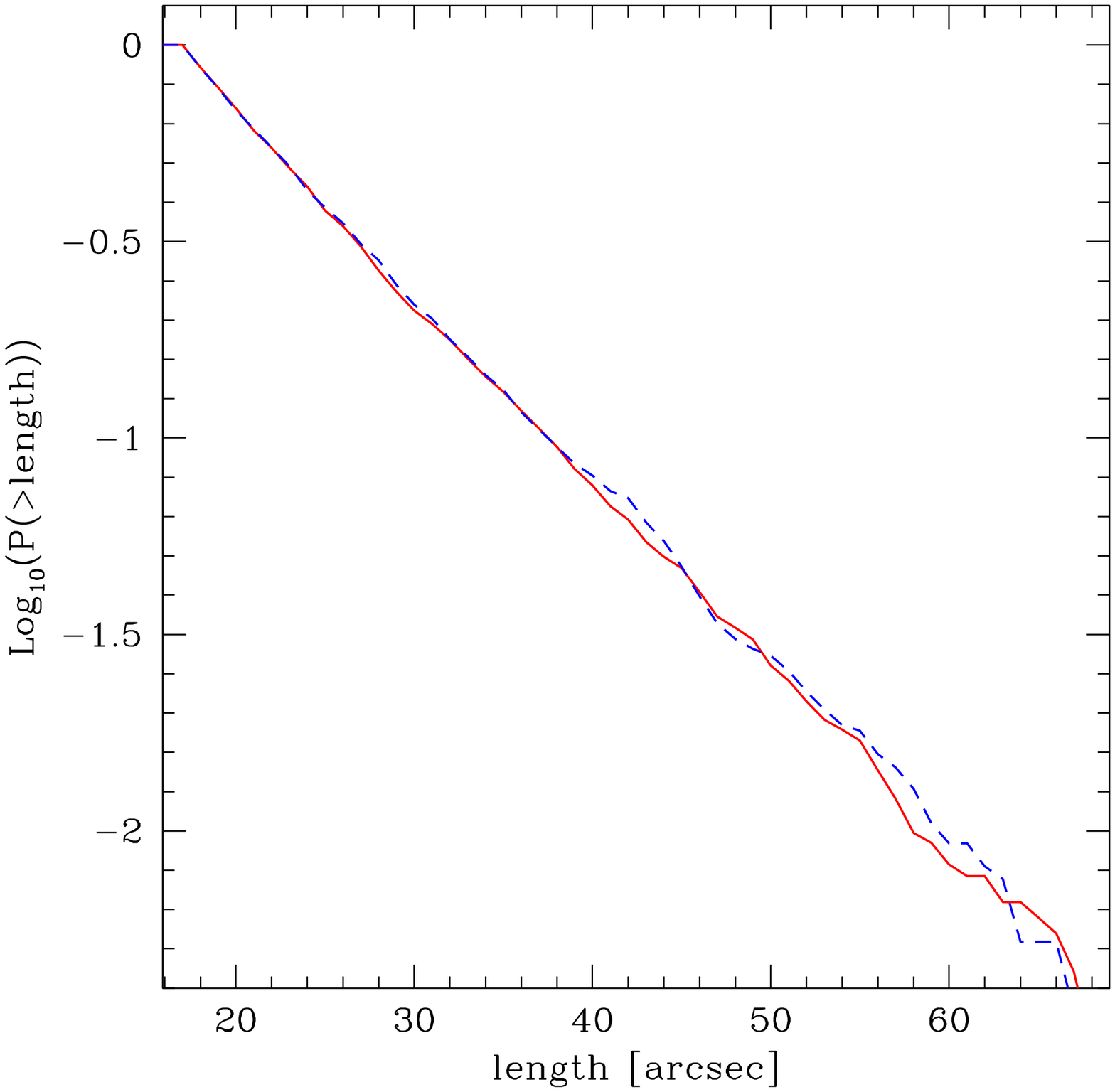,height=2.05in} 
\psfig{figure=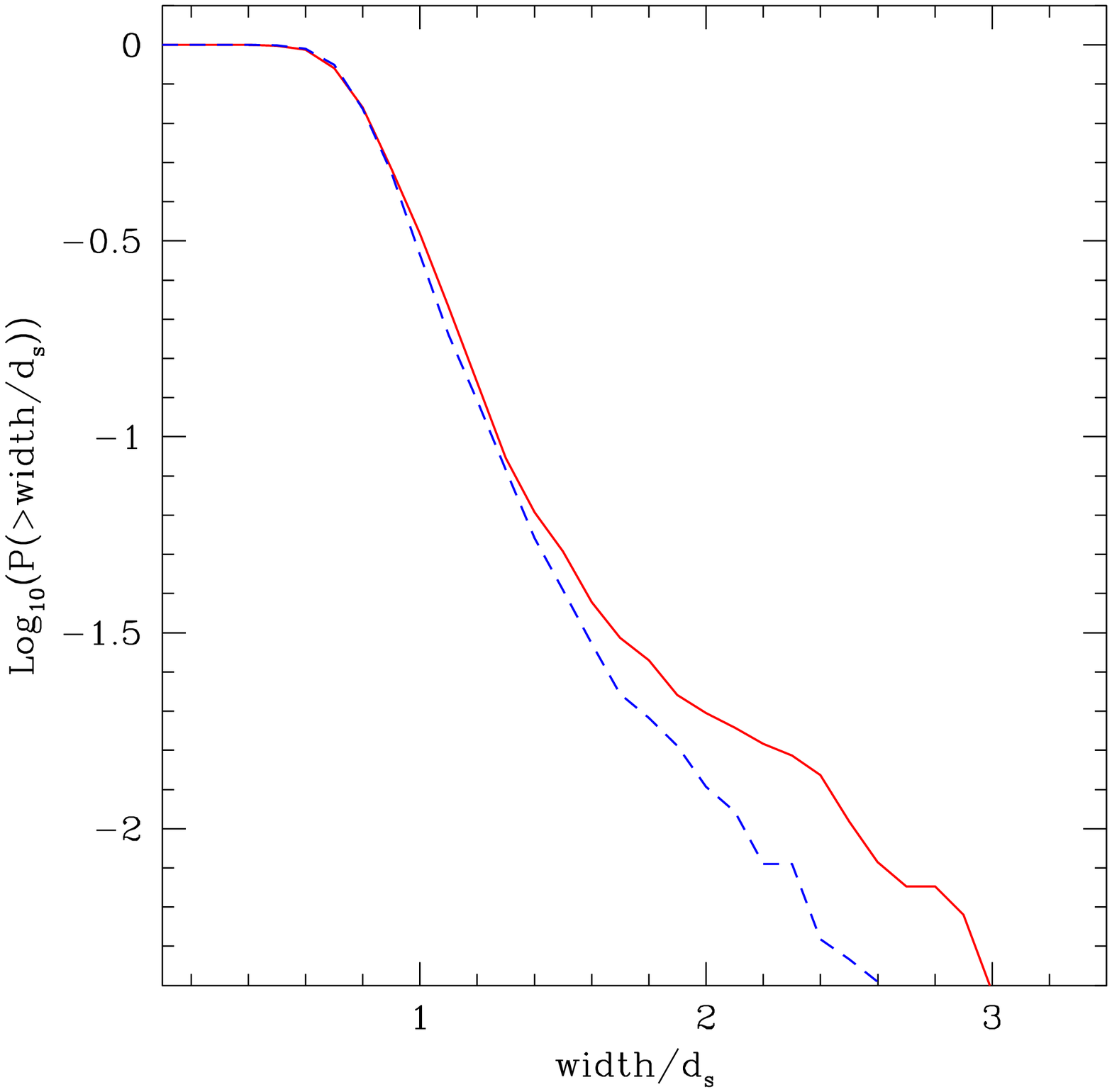,height=2.05in} 
\psfig{figure=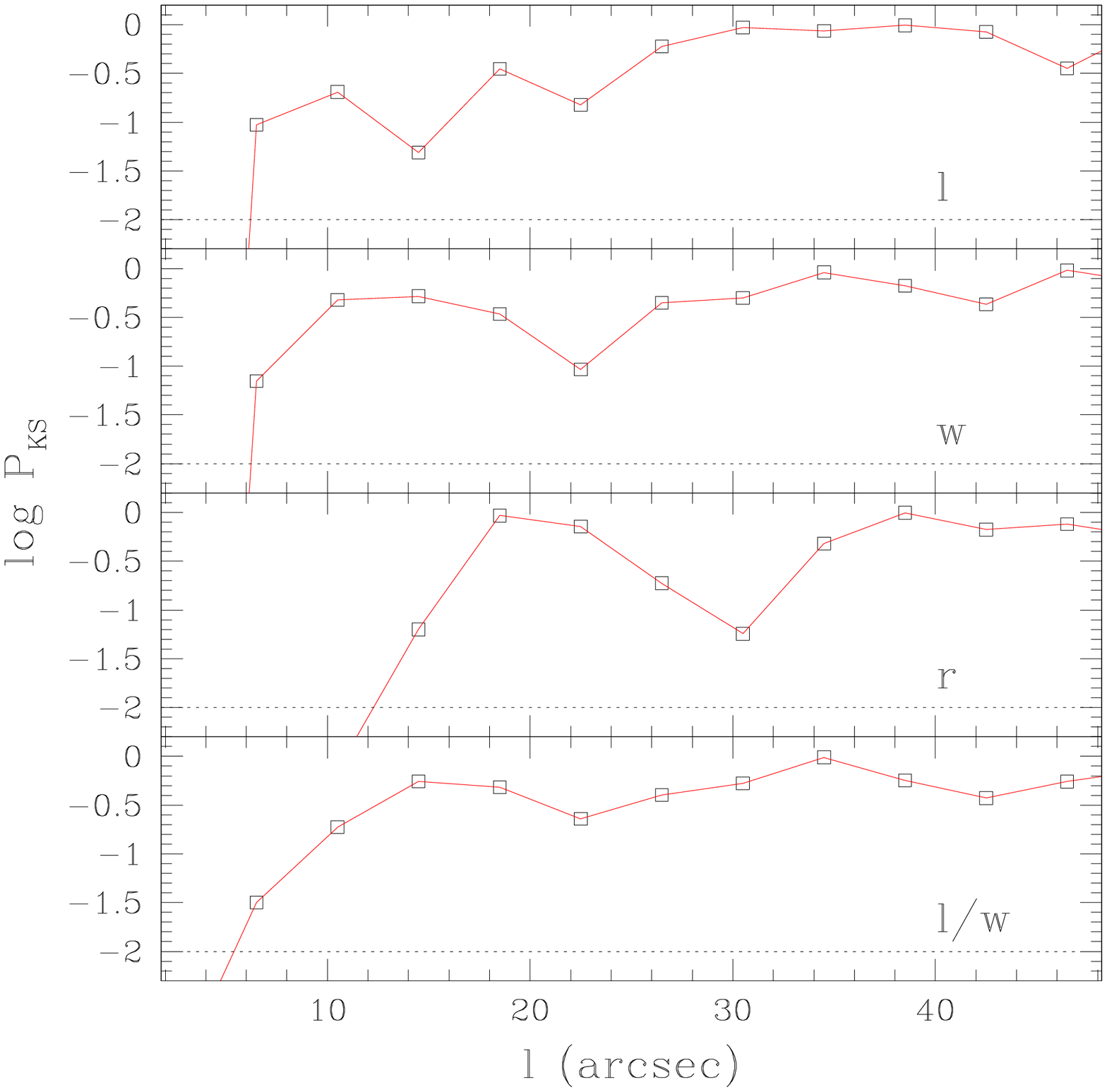,height=2.05in}} 
\caption{Cumulative distributions of arc 
lengths (left panel) and widths (central panel) for arcs longer than 
$16''$. Results for DM and GAL simulations are shown by solid and 
dashed lines, respectively. Right panel: the behaviour of the 
probability $P_{KS}$ (as computed in a Kolmogorov-Smirnov test) that 
the arc property distributions in data sets obtained from the 
simulations DM and GAL can be drawn from the same parent 
distribution. Subsamples of arcs with a given length $l \pm 2''$ are 
considered. The subpanels refer to different properties: length $l$, 
width $w$, curvature radius $r$ and length-to-width $l/w$ from top to 
bottom. 
\label{cumulatives}} 
\end{figure} 
 
\section{Properties of arcs and results} 
We use our clusters to lens a large number of elliptical sources (with 
axial ratios randomly drawn in the interval [0.5,1] and area equal to 
that of a circle of diameter $2''$) on the source plane. Following the 
method introduced by Miralda-Escud\'e \cite{miralda93b} and developed 
by Bartelmann \& Weiss \cite{bartelmann94}, we find and classify the 
images of all these sources, measuring their lengths $l$, widths $w$, 
curvature radii $r$ and length to width ratios $l/w$. We then perform 
a statistical analysis of the distributions of the arc properties 
(more details are presented in \cite{meneghetti00}). 
   
The results obtained from the first set of 27 simulations using the 
original simulated clusters (i.e. without galaxies inside, hereafter 
``DM'' simulations) are compared to those obtained after the 
introduction of the galaxies in the lens clusters (hereafter ``GAL'' 
simulations). 
 
The total number of arcs in the DM and GAL simulations is quite 
similar: 447112 and 448927 respectively. The majority of these arcs 
are quite short. Considering only giant arcs with $l>16''$ the sample 
reduces to 1823 and 1721 arcs for DM and GAL simulations.  In Figure 
\ref{cumulatives} we show the cumulative distributions of lengths and 
widths for arcs longer than $16''$.  The distributions of arc length 
do not seem to be sensitive to the inclusion of the galaxies in the 
clusters.  We found a similar result also for the distributions of arc 
curvature radii. On the other hand, the distributions of arc widths 
show some differences between simulations DM and GAL. Such differences 
are partially found also between the distributions of arc length to 
width ratios. 
 
Considering arcs with a given length $l \pm 2''$, we perform the 
Kolmogorov-Smirnov test to evaluate the significance of the 
differences between the distributions of the arc properties. The 
significance level obtained from the test as a function of $l_{min}$ 
is also shown in Figure \ref{cumulatives} for all the considered arc 
properties. Concerning arc widths, the probability that the data sets 
obtained from simulations DM and GAL are drawn from the same 
distribution becomes slightly lower for arcs with $l< 22''$. On the 
other hand, the differences between the other property distributions 
become significant only when very short arcs are included in the 
sample. 
 
\begin{figure} 
\centering 
{\psfig{figure=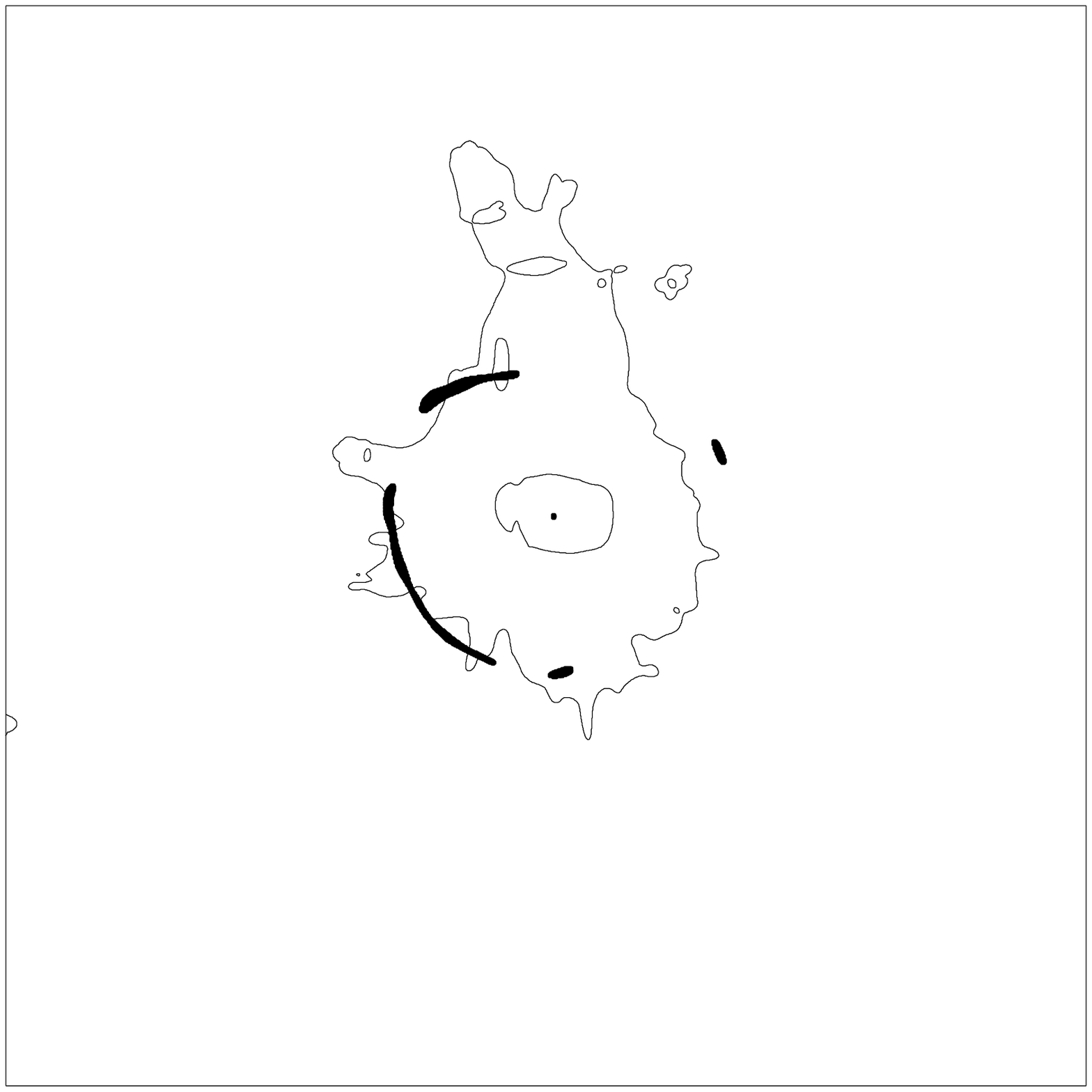,height=2.4in} 
\psfig{figure=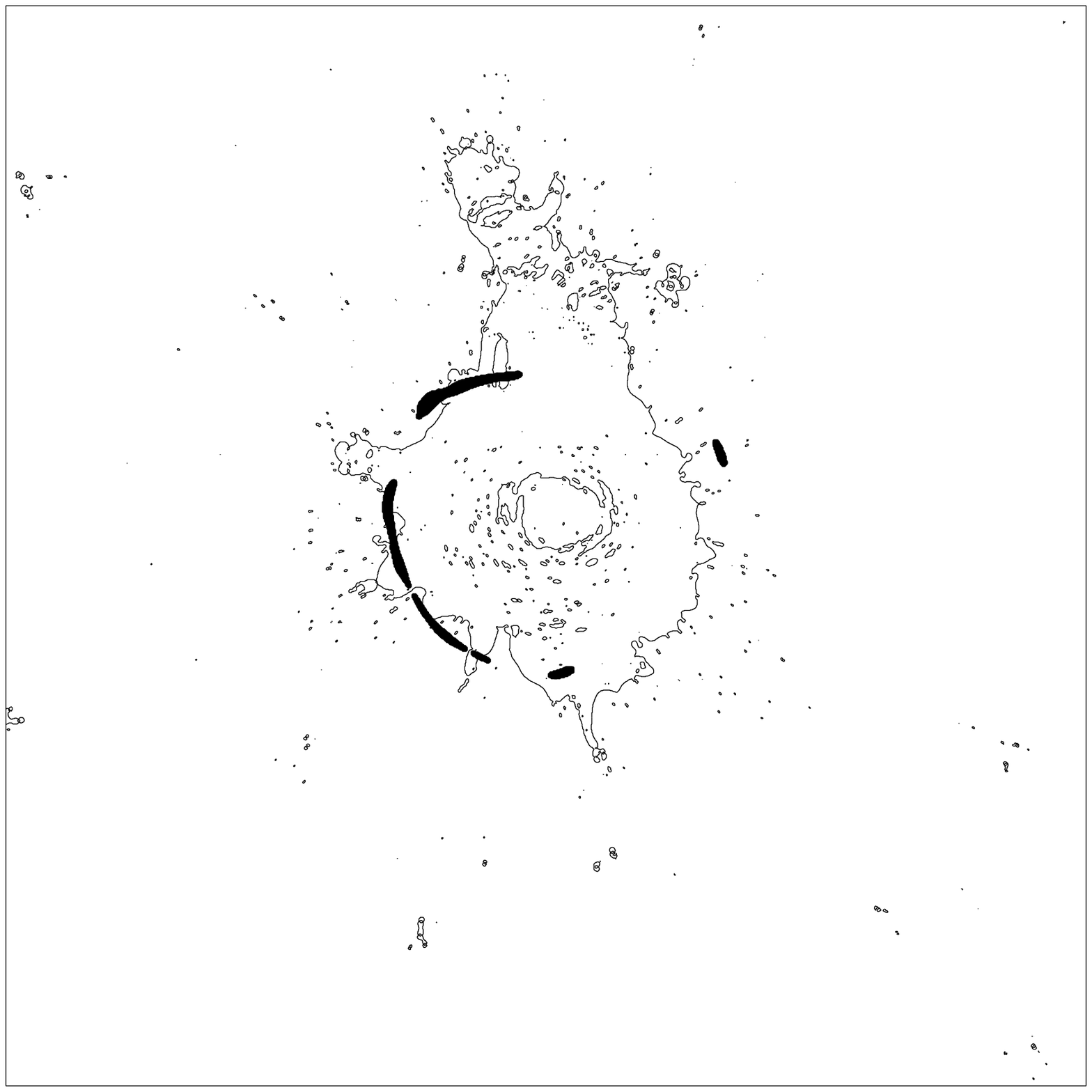,height=2.4in}} 
\caption{Example of arc perturbed and splitted by cluster 
galaxies. The left panel refers to the DM simulation, while the right 
one shows the image of the same source in the corresponding GAL 
simulation. We show also critical curves, which tend to be wiggled 
when the galaxies are present. 
\label{arcs}} 
\end{figure} 
 
\section{Conclusions} 
We expected these effects by cluster galaxies: {\em a)} they tend to 
increase the cluster cross section for strong lensing by wiggling the 
cluster critical curves, increasing their length; {\em b)} because of 
the larger curvature of the critical lines, they can perturb arcs, 
splitting them; {\em c)} the local steepening of the density profile 
near cluster galaxies tends to make arcs thinner. 
 
The results of the KS test indicate that, concerning the arc lengths, 
curvature radii and length to width ratios, the effect of cluster 
galaxies on arc statistics is negligible if very short arcs are 
excluded. This means that the first two effects previously mentioned 
are almost exactly counter-acting and that the splitting of some arcs 
is partially compensated for the increased strong lensing ability of 
the clusters (Figure \ref{arcs}).  The results show also that, as 
expected, the galaxies tend to make arcs thinner. This effect becomes 
evident for arcs shorter than $22''$. 
 
The longest arcs, which form in the central regions 
of the clusters, where most of the mass is concentrated, are not 
sensitive to cluster galaxies. We think that this is due to the fact 
that in these dense regions only a small fraction of the total galaxy 
mass emerges from the underlying dark matter distribution.

\end{document}